\documentclass[11pt]{article}
\AtBeginDocument{%
  }

\usepackage[a4paper,margin=1in]{geometry}
\usepackage[T1]{fontenc}
\usepackage[utf8]{inputenc}
\usepackage[blocks]{authblk}
\usepackage{lmodern}
\usepackage{microtype}
\usepackage{booktabs}
\usepackage{tabularx}
\usepackage{array}
\usepackage[colorlinks=true]{hyperref}
\usepackage{xurl}
\usepackage{enumitem}
\usepackage{stfloats}
\usepackage{multirow}

\usepackage{listings}
\usepackage{xcolor}

\usepackage{tikz}
\usepackage{orcidlink}

\lstdefinelanguage{TOML}{
  morecomment=[l]{\#},
  morestring=[b]",
  alsoletter={._-},
  morekeywords={image,mounts,workdir,entrypoint,devices,env,annotations},
  sensitive=true
}

\title{Sarus Suite: Cloud-native Containers for HPC}

\author[1]{Alberto Madonna\orcidlink{0009-0005-5844-5488}}
\author[1]{Matteo Chesi\orcidlink{0000-0002-1204-7565}}
\author[1]{Gwangmu Lee\orcidlink{0009-0006-6464-355X}}
\author[1]{Michele Brambilla\orcidlink{0000-0002-5547-4516}}
\author[1]{\\Fawzi Roberto Mohamed\orcidlink{0009-0003-8907-4427}}
\author[1]{Felipe A. Cruz\orcidlink{0009-0002-0921-6134}}

\affil[1]{ETH Zurich / Swiss National Supercomputing Centre, Lugano, Switzerland}
\affil[ ]{\texttt{\{alberto.madonna, matteo.chesi, gwangmu.lee, michele.brambilla, fawzi.mohamed, felipe.cruz\}@cscs.ch}}

\date{} 

\begin{document}

\maketitle

\begin{abstract}

High-performance computing (HPC) systems must support fast-moving software stacks, especially in AI/ML, while preserving scheduler control, scalable startup, and production performance. Yet many HPC container solutions rely on specialized runtime stacks that weaken continuity with mainstream cloud-native workflows and require ongoing effort to sustain compatibility with the evolving upstream ecosystem.
We argue that HPC should specialize the integration layer while keeping the container engine aligned with upstream container evolution.

We present \emph{Sarus Suite}, an upstream-aligned HPC container architecture built around an unchanged Podman engine. Sarus Suite adds the HPC-specific functionality needed for production use through complementary system layers for declarative runtime specification, scheduler-native execution, scalable shared-image access, and standards-based host capability injection.

We evaluate Sarus Suite on a Cray EX GH200 system using communication-intensive HPC workloads, large scale AI training, metadata-heavy startup workloads, and container startup measurements. Across PyFR, SPH-EXA, Megatron-LM, and Pynamic, Sarus Suite matches the performance and scaling of the production Enroot+Pyxis baseline while delivering consistently faster per-node container startup. The architecture also enables direct use of upstream OCI images, including NGC-based images, and supports cloud-native multi-container workflows expressed through Kubernetes manifests.

These results show that HPC-grade containers do not require an HPC-specific runtime, provided that scheduler semantics, scalable image access, and host integration are implemented in explicit system layers. This preserves upstream continuity and software agility while maintaining scheduler control, scalability, and production performance.

\end{abstract}

\section{Motivation}

High-performance computing platforms increasingly face a software-change problem. Production systems are expected to remain stable, predictable, and supportable over multi-year lifecycles, while the software ecosystems most relevant to contemporary AI/ML and data-intensive science evolve on much shorter timescales. Frameworks, toolchains, and dependency stacks now change far more quickly than the platforms on which they must run, creating a persistent tension between operational stability and software agility. In this setting, software deployment is no longer merely a matter of user convenience or packaging efficiency; it becomes a first-order systems concern, because unmanaged software change can propagate operational risk across shared infrastructure.

Container images compatible with cloud-native~\cite{cloud-native-definition} and Open Container Initiative (OCI) standards~\cite{oci-image-spec,oci-runtime-spec,oci-distribution-spec} are attractive in this setting because they package software environments as versioned, externally maintained artifacts that can move across development, testing, and production. Rather than rebuilding and revalidating the same stack independently at each site, users and operators can rely on standard images built upstream, distributed through registries such as NVIDIA NGC, and promoted across stages of the software lifecycle with much less translation. This is especially valuable for fast-moving software ecosystems, where treating the environment as a stable artifact shortens validation cycles, improves reproducibility, and reduces the coordination burden associated with frequent updates.

For HPC, however, image portability alone is not enough. What increasingly matters is continuity with the broader container ecosystem built around OCI standards: engine behavior, registry workflows, build pipelines, lifecycle semantics, extension interfaces, and higher-level tooling. Modern software practice depends not only on the portability of the image format, but also on the workflows and interfaces through which environments are developed, tested, distributed, and executed. If execution on HPC depends on a specialized runtime model that diverges from those interfaces and assumptions, then the image may travel while the surrounding workflow does not. In that case, HPC continues to lag behind mainstream container practice and must repeatedly recover new capabilities through local reinvention rather than inheriting them through upstream evolution.

HPC containerization has matured into a standard part of production system design, and prior solutions have shown that scheduler-aware execution, efficient startup, controlled host integration, and scalable application performance are all achievable in practice. The central question is therefore no longer whether containers can be used effectively in HPC, but what container architecture best preserves both HPC operational requirements and alignment with the broader software landscape.

The problem is thus architectural. HPC clearly needs specialized mechanisms around containers, including scheduler control, scalable image access, policy enforcement, and host capability injection. The question is where that specialization should be located. Our position is that HPC should specialize system integration while preserving mainstream container semantics and interfaces wherever possible. In this model, effort is concentrated on the parts that are genuinely site- and system-specific, while the main container engine remains aligned with the broader cloud-native ecosystem.

From this architectural position, prior HPC container solutions can be understood as establishing key parts of the required mechanisms. Enroot/Pyxis demonstrates scheduler-centric one-container-per-node execution with efficient squashed-image startup; the original Sarus engine~\cite{sarus-2019} demonstrates OCI-oriented host integration; and podman-hpc shows that Podman can be adapted to HPC environments. Collectively, these tools prove that the relevant mechanisms are already available in separate forms. What remains missing, to our knowledge, is to combine and evaluate these mechanisms as a single architecture built around an unchanged upstream Podman engine, with HPC semantics introduced only through supported extension and configuration interfaces. Section~\ref{sec:related-work} discusses related HPC container implementations in more detail.

Sarus Suite~\cite{sarus-suite-software} addresses this gap. It builds around Podman~\cite{podman-software}, used unchanged as the engine, and provides the missing HPC integration through scheduler-aware execution, declarative environment specification, standards-based host injection, and scalable image distribution. Sarus Suite is thus best understood not as a new container engine, but as an architectural approach for combining production HPC requirements with continued alignment to the broader and continuously evolving cloud-native container ecosystem.
This paper makes four main contributions. First, it presents an upstream-aligned architecture for HPC containers that keeps a full mainstream OCI engine as the user-facing interface and introduces HPC-specific elements only through supported extension and configuration mechanisms. Second, it develops a scheduler-controlled execution model that preserves Slurm semantics, avoids per-rank container instantiation, and integrates containerized execution into the workload manager’s normal lifecycle. Third, it combines a declarative environment model with a scalable image-access path that reduces launcher complexity and launch-time filesystem pressure while preserving standard container images and engine workflows. Fourth, it evaluates the resulting architecture on a Cray EX GH200 system, demonstrating scalable startup and competitive performance for representative HPC and AI workloads.

\section{Designing Cloud-Native HPC Containers}

Taking the architectural position that an HPC container-stack should specialize system integration while preserving mainstream container semantics and interfaces wherever possible, this section discusses the resulting container-stack design. The key question is not whether OCI images can package software successfully, but what execution architecture allows portable software environments to be separated from system-specific HPC instantiation.

For HPC, the main value of cloud-native containers is that they separate a portable software environment from the system-specific mechanisms required to execute that environment correctly on a particular machine. In this view, the image defines the application environment, while the container stack remains responsible for instantiating it under the constraints of the target system.

That container-stack responsibility includes HPC site-specific details on which correct execution depends: accelerators, communication libraries, GPU drivers, PMI integration, and other resources whose compatibility and policy must be dictated by the system. These are not just convenience features; they are part of the platform contract required to run efficiently and correctly on production infrastructure.

Embedding such assumptions directly into images improves local convenience at the cost of portability and maintainability. Images that carry site-local MPI stacks, interconnect-specific libraries, or environment-specific launcher logic may work on one system, but they become brittle artifacts whose correctness depends on assumptions that do not transfer cleanly across platforms or future stack revisions. Their use also discourages access to the broader OCI ecosystem, including externally maintained images such as those distributed through NVIDIA NGC.

The architectural consequence is therefore straightforward: keep images portable and versioned, and move runtime specialization into explicit interfaces controlled by the platform. This preserves the value of standard OCI images while allowing the HPC system to provide the host integration and policy enforcement required for production execution.

Once portable images and system-controlled instantiation are separated, four operational requirements follow.

First, container execution must remain under workload-manager control. Scheduling, allocation, accounting, and process-lifecycle semantics are core properties of HPC operation and user experience, so containerized jobs must remain aligned with the same scheduler model that governs native execution.

Second, image access must scale under synchronized launch. Frequent pull-and-unpack behavior across many nodes creates avoidable I/O overhead and can be infeasible on diskless compute nodes. Even with shared storage, naive use of unpacked OCI layers can generate substantial metadata pressure during large launches.

Third, access to custom host capabilities must be injected at run time under container-stack control rather than embedded permanently into the image. This includes accelerators, proprietary communication libraries, driver-dependent components, and other system-tuned runtime stacks whose correctness and performance depend on the target machine.

Fourth, the intended execution environment should be expressed declaratively. Without such a mechanism, users reconstruct deployment conditions repeatedly through launcher glue in job scripts, making behavior harder to validate, reproduce, and share across workflows and sites.

These requirements do not imply a new container abstraction. They imply an architecture around standard OCI artifacts in which HPC-specific behavior is added through system integration rather than by changing the container interface and semantics exposed to users.

In fact, replacing mainstream container registry workflows, build pipelines, and lifecycle semantics with an HPC-specific runtime would duplicate effort around functionality that is already standardized and evolving upstream and create an ongoing maintenance burden.

Given these requirements, the architectural task is therefore to retain a full, standard OCI engine for image and container semantics while placing HPC-specific behavior in complementary integration components.

Podman fits this role well. It provides mainstream OCI-compatible image and container management, supports daemonless and rootless operation suited to multi-user systems, offers layered configuration for combining site defaults with user settings, and exposes standard extension interfaces such as OCI hooks and CDI. Just as importantly, it does not impose cluster-orchestration assumptions, which allows scheduler integration and performance policy to remain external and site-controlled rather than hard-wired into the engine.

The resulting separation of concerns is clear. The engine remains responsible for baseline OCI semantics and image handling; complementary system components provide scheduler-aware execution, scalable image access, controlled host-capability injection, and declarative runtime specification. This keeps HPC systems close to upstream container practice while specializing only the parts that are genuinely HPC system-specific. Sarus Suite adopts exactly this design stance.

\section{Sarus Suite Execution Model}

\begin{figure*}
 \centering \includegraphics[width=\textwidth]{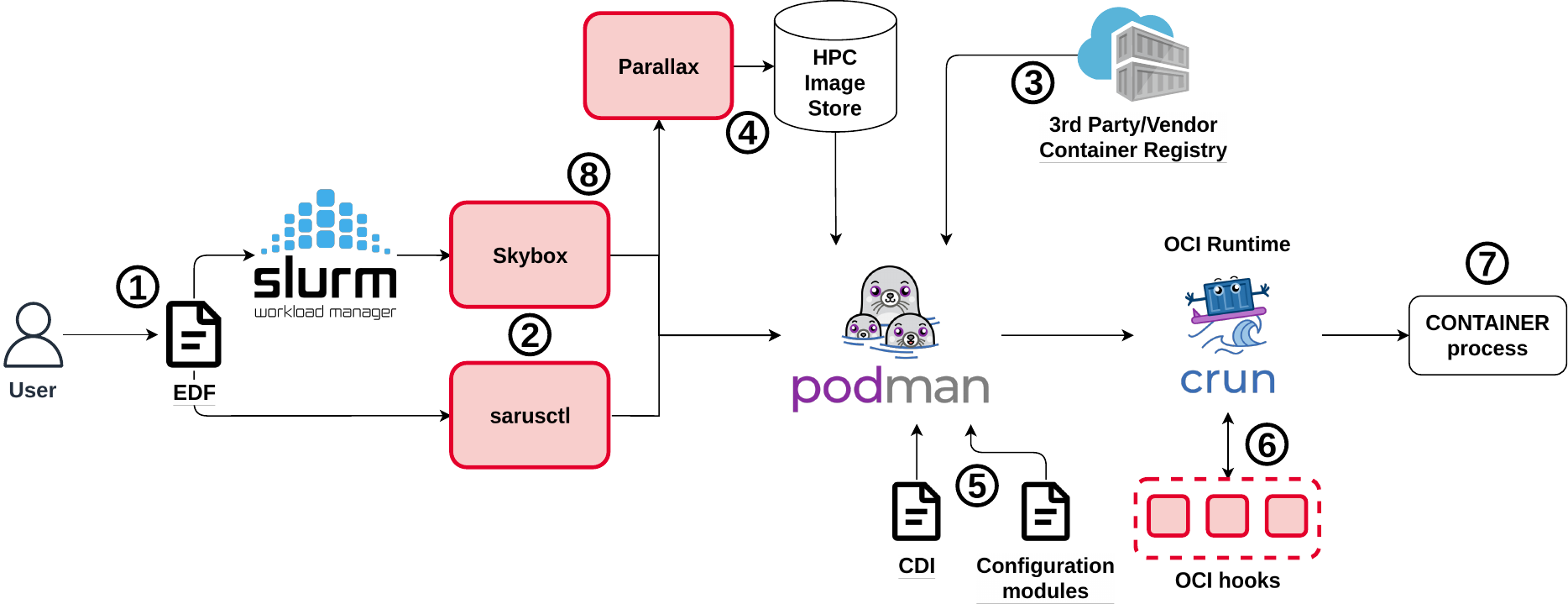}
 \caption{Sarus Suite execution flow: (1) the user describes the container environment in the EDF; (2) the workload is launched via \texttt{sarusctl} or the Skybox Slurm plugin; (3) Podman retrieves the image from a 3rd party registry if needed; (4) Parallax stages the image in the HPC image store for scalable access; (5) Podman creates the container using CDI and site-controlled configuration; (6) OCI hooks inject host libraries, devices, and other resources; (7) the OCI runtime starts the container process; and (8) Skybox applies scheduler semantics, including one-container-per-node execution, rank placement, and cleanup.}
 \label{fig:concept-arch}
 \end{figure*}

Sarus Suite implements the architectural position developed earlier through four functionalities: declarative runtime specification, scalable image access, controlled host-capability injection, and scheduler-aware execution. Together, these functions turn a portable container image into a scheduler-managed, system-integrated execution environment suitable for HPC. Rather than asking users to reconstruct that environment through per-job launcher logic, Sarus Suite separates portable application content from the system-specific decisions required to instantiate it correctly and efficiently on the target machine.

The execution model is organized around a small set of components with distinct roles. The Environment Definition File (EDF) records the intended runtime environment as a reusable object. Podman remains the OCI engine for standard image and container operations. The Parallax storage utility provides an image-access path suited to shared HPC storage. CDI resource descriptions, OCI hooks, and site-controlled configuration supply launch-time host integration. The Skybox plugin embeds containerized execution within Slurm's control model. Figure~\ref{fig:concept-arch} summarizes how these components cooperate during execution.

\paragraph{Declarative runtime specification.}

Sarus Suite begins from the premise that the intended execution environment should be expressed once as a reusable runtime object rather than reconstructed repeatedly through job-script glue. That role is served by the EDF, which records the image to be used together with the runtime configuration and requested HPC integrations needed to execute it on the target system. In a typical workflow, users select or build a container image, validate and publish it through ordinary CI/CD or registry workflows, author an EDF describing the intended runtime environment, and then submit through Slurm.
This separation decouples portable image construction from site-specific execution. The image remains a standard OCI artifact that can be built, tested, versioned, and distributed through mainstream container workflows, while the EDF captures the runtime intent that must later be interpreted under platform control. The EDF is therefore the transition point at which a portable software environment becomes a schedulable HPC execution object: it specifies \emph{what} environment is required without forcing users to encode \emph{how} that environment must be instantiated on a particular system.

\paragraph{Scalable image access.}

After runtime intent has been declared, the referenced image must be made available in a form suitable for large-scale execution. Podman preserves standard OCI workflows for image retrieval and baseline container operations, while Parallax provides an image-access designed for parallel multi-node startup on shared HPC storage.
Standard OCI layer handling is not well suited for that setting, where repeated pull-and-unpack behavior can introduce avoidable startup overhead and substantial metadata pressure on parallel filesystems. Parallax transforms retrieved images into a shared runtime representation designed for efficient reuse across the filesystem visible to the allocated nodes.

\paragraph{Controlled host-capability injection.}

Portable images must be complemented at launch time with the host-specific resources required for correct and efficient HPC execution. Sarus Suite therefore keeps devices, communication libraries, host drivers, and related system-tuned runtime components under platform control rather than embedding them permanently into the image.
This is implemented through a combination of site-controlled configuration, CDI resource descriptions, and OCI hooks. Site configuration establishes baseline HPC container settings. CDI describes how devices and other managed resources are exposed to the container. OCI hooks apply the runtime modifications needed to make host libraries, communication stacks, and related software components available inside the instantiated environment. Together, these mechanisms let the container-stack inject the resources required by the workload while keeping those decisions external to the portable container image itself.

\paragraph{Scheduler-aware execution.}

Containerized execution must remain within the workload manager's control model. Sarus Suite implements this capability through Skybox.

For Slurm jobs, Skybox creates one container per allocated node and then places ranks into the appropriate namespaces, preserving the expected relationship between allocation, launch, and process placement. Containerized workloads therefore remain governed by the same scheduler semantics as native jobs: allocation ownership and rank placement stay with Slurm, accounting follows the normal job model, and cleanup remains tied to scheduler-controlled process lifecycle.

The four functions outline in this Section establish the execution responsibilities that the Sarus Suite stack must realize. The next section turns from this functional architecture to the concrete implementation, detailing how the corresponding components, interfaces, and integration are implemented in the system.

\section{Implementation components}\label{sec:components}
Sarus Suite is a small, composable set of utilities that make HPC container environments portable, policy-compliant, and scalable. The design separates \textit{what} an environment is, \textit{how} it integrates with the host, and \textit{how} it is executed and distributed.

\subsection{EDF: Declaring Runtime Intent}
\label{subsec:edf}

The Environment Definition File (EDF) makes the execution environment as a whole, rather than the image alone, the primary reusable object under a Sarus Suite workflow. An EDF is a declarative specification (e.g., TOML) evaluated under workload-manager control at job start, after the image has been built and validated but before the workload is instantiated on the system. In this way, the EDF decouples three concerns that are often tangled in practice: the container image as a portable software artifact, the runtime configuration needed for a specific workload, and the launch mechanism through which that workload is submitted. The image remains portable and externally maintainable, while the EDF captures the intended runtime environment as a versioned, shareable object that can be re-instantiated across runs and sites.

An EDF records standard runtime choices such as the OCI image reference, bind mounts, working directory, entrypoint behavior, and environment variables. More importantly for HPC, it also provides an structured interface for requesting optional system integrations at launch time. In practice, EDF annotations act as opt-in integration switches that express \emph{intent} rather than prescribing a fixed implementation. An annotation therefore requests a capability (for example, communication support, GPU-related runtime services, or framework-specific adaptation), while the concrete resolution of that request is delegated to the underlying site- and system-specific container setup. This keeps host- and system-specific behavior out of the EDF and out of the image itself: the same EDF can request the same capability across systems, while the local runtime stack determines how that capability is provided on a given platform.

The effect is to turn containerized execution into a stable runtime contract rather than a collection of per-job flags. Instead of repeatedly reconstructing mounts, environment propagation, cache paths, device exposure, and integration hooks in Slurm job scripts, the user references a named environment and Sarus Suite resolves the corresponding image access, runtime policy, and host capability injection underneath. The EDF therefore gives Sarus Suite a concrete mechanism for separating portable workload intent from site-controlled actualization: users declare the environment they need, while the system determines the correct implementation for the allocated resources and platform policy.

\begin{figure}[t]
\begin{minipage}{\linewidth}
\begin{lstlisting}[caption={EDF fragment illustrating portable runtime configuration. HPC integrations are requested via annotations, and CDI specs activated through the devices array.},label={lst:edf-example}]
image = "ghcr.io/cscs/ml-workflows/transformers:latest-arm64"

mounts = [
  "/scratch/$USER/hf-models:/opt/hf",
  "/scratch/$USER/data:/data",
  "/scratch/$USER/output:/output"
]

workdir = "/scratch/$USER"
entrypoint = false

devices = ["nvidia.com/gpu=all"]

[env]
HUGGINGFACE_HUB_CACHE = "/opt/hf/hub"
TRANSFORMERS_CACHE    = "/opt/hf/transformers"

[annotations]
com.hooks.cxi.enabled             = "true"
com.hooks.aws_ofi_nccl.enabled    = "true"
com.hooks.aws_ofi_nccl.variant    = "cuda12"
com.hooks.nvidia_cuda_mps.enabled = "true"
\end{lstlisting}
\end{minipage}
\end{figure}

\begin{figure}[t]
\begin{minipage}{\linewidth}
\begin{lstlisting}[caption={\texttt{podman run} expansion of Listing 1's EDF, demonstrating Sarus Suite's automatic orchestration of low-level Podman options from the declarative configuration.},label={lst:cli-example}]
podman \
  --ipc host --network host --pid host --uts host \
  --userns keep-id --cgroupns host --cgroups no-conmon --tz local \
  --root "$PODMAN_ROOT" \
  --runroot "$PODMAN_RUNROOT" \
  --storage-opt "additionalimagestore=$RO_STORAGE" \
  --storage-opt "mount_program=/usr/bin/parallax-mount-program" \
  run \
  --mount "type=bind,src=/scratch/$USER/hf-models,dst=/opt/hf" \
  --mount "type=bind,src=/scratch/$USER/data,dst=/data" \
  --mount "type=bind,src=/scratch/$USER/output,dst=/output" \
  --workdir "/scratch/$USER" \
  --entrypoint "" \
  --device "nvidia.com/gpu=all" \
  --env "HUGGINGFACE_HUB_CACHE=/opt/hf/hub" \
  --env "TRANSFORMERS_CACHE=/opt/hf/transformers" \
  --annotation "com.hooks.cxi.enabled=true" \
  --annotation "com.hooks.aws_ofi_nccl.enabled=true" \
  --annotation "com.hooks.aws_ofi_nccl.variant=cuda12" \
  --annotation "com.hooks.nvidia_cuda_mps.enabled=true" \
  ghcr.io/cscs/ml-workflows/transformers:latest-arm64
\end{lstlisting}
\end{minipage}
\end{figure}

Listing~\ref{lst:edf-example} illustrates how the EDF combines common container runtime settings with explicit requests for HPC-specific launch-time integrations, allowing the image to remain portable while site-controlled extensions are applied only at instantiation time. In this sense, EDF annotations are portable capability requests: they standardize the \emph{what} of integration, while leaving the \emph{how} to the local runtime, engine configuration, and site policy. Listing~\ref{lst:cli-example} makes the complementary point that the equivalent \texttt{podman run} command is substantially more verbose because it must explicitely setup the low-level engine, namespace, storage, mount, device, and runtime options required on the target platform. Taken together, the two listings show how EDF both reduces per-job launcher complexity and abstracts over system-binding details that would otherwise be embedded in job scripts or custom site-specific launch logic.

\subsection{Sarusctl and Skybox}
Sarus Suite provides end-to-end tooling so the EDF you validate is the one that runs. In practical use, this separation also follows the expected preparation workflow. Users can validate an EDF interactively with \texttt{sarusctl} and confirm that the requested image and launch-time integrations behave as intended. This in turn migrates the image into the Parallax shared store before submitting the large Slurm job. As a result, the first-use image acquisition path is supported by Skybox but is not the representative execution mode for production jobs. The common case for scaled execution is instead that the EDF and image have already been exercised at small scale with \texttt{sarusctl} and the Parallax artifact is already present in the shared store, so the large launch follows the warm-start path. \texttt{sarusctl} is a Rust CLI that parses and validates EDFs, manages images via Podman+Parallax (migrate/query/remove), and runs EDF-described containers interactively for pre-flight testing. Skybox reuses the same parsing and configuration logic inside the workload manager, avoiding drift between interactive and batch execution. In Slurm, Skybox implements this as a SPANK plugin that creates containers during \texttt{task\_init}, runs one container per node, moves ranks into container namespaces via \texttt{setns}, and cleans up in \texttt{task\_exit}.

\subsubsection{Skybox: Slurm integration}
\label{subsubsec:skybox-slurm-integration}
Skybox is implemented as a Slurm SPANK plugin that embeds containerization into Slurm's normal job-step execution model rather than launching each rank through an independent container runtime process. In the local and allocator SPANK contexts, Skybox registers the \texttt{-{}-edf} option, renders the EDF once, and serializes the expanded result into the job environment. In the remote context, it reconstructs the EDF from that serialized representation.

\begin{figure*}[t]
  \centering
  \includegraphics[width=\textwidth]{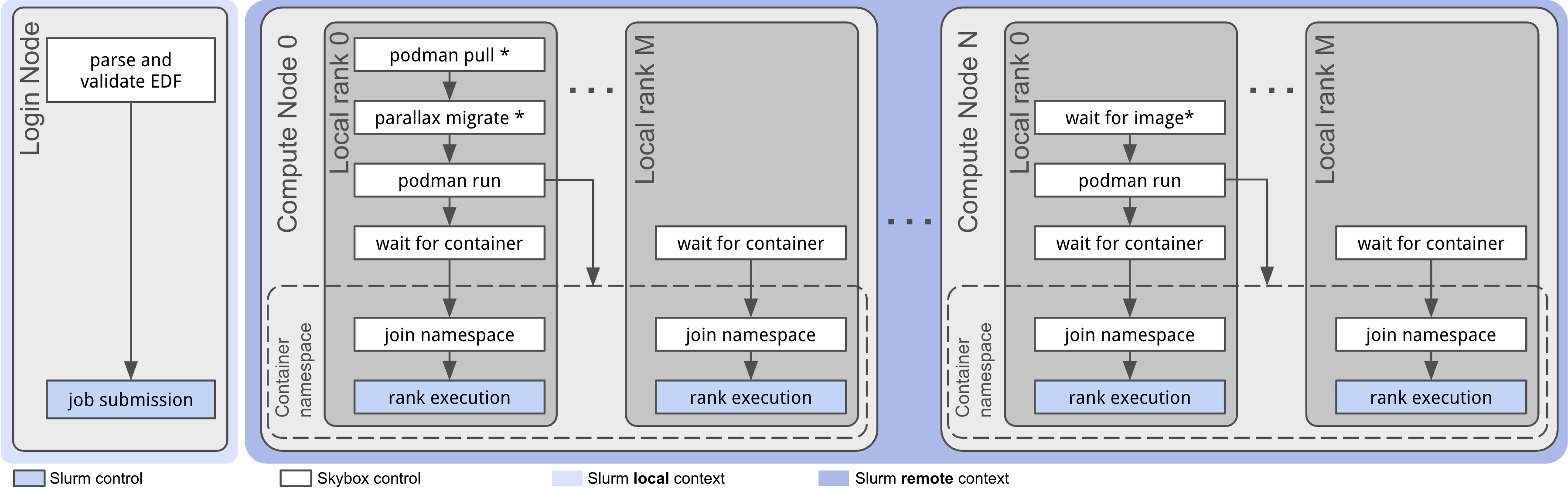}
  \caption{Skybox execution model under Slurm. The EDF is rendered once and distributed as a runtime object, with optional  image acquisition via Parallax. On each node, local task~0 launches the container via Podman; ranks then join its namespaces before execution proceeds. The figure highlights the separation between job-wide and rank-local operations.}
  \label{fig:skybox-slurm-flow}
\end{figure*}

\paragraph{Execution model.}
Figure~\ref{fig:skybox-slurm-flow} summarizes the Slurm-integration execution model.
Skybox uses a one-image-per-job, one-container-per-node execution strategy.
This design targets HPC workloads with many ranks per node, where launching every rank through an independent container-runtime process can make startup costs scale with rank count and can interfere with host-integrated transports and process coordination.
By managing containerization as a node-level setup, Skybox reduces launch times and improves scalability for many-ranks-per-node workloads.
Across the entire job, only \textit{global} task~0 performs image acquisition when needed.
On each node, only \textit{local} task~0 starts a detached container through Podman.
This node-individual container is not used to launch the application directly; instead, at this stage the only process present inside the container is a lightweight \textit{shim process}, whose only purpose is to keep the container namespaces alive, providing the container filesystem view and user-namespace context for the duration of the job step.
Once the container is ready, Skybox drives each rank to enter its namespaces and yields control back to Slurm, which proceeds to launch user commands normally, but now transparently located inside the container.

This is central to Skybox's scalability. The image handling happens once for the whole job, container startup is done once per node, and rank creation remains under Slurm control. Skybox therefore avoids extra per-rank container instantiation and resource utilization while preserving normal Slurm ownership of rank placement, signaling, accounting, and teardown. This is valuable not only because it preserves scheduler compatibility, but because it retains the native HPC division of responsibilities for containerized jobs. Users still describe resources and the intended runtime environment, while Slurm remains responsible for rank lifecycle and resource control. By moving runtime intent into the EDF and keeping execution under Slurm ownership, Sarus Suite avoids reconstructing mounts, environment propagation, device exposure, and integration hooks in application-side logic. This closes the user-experience gap between native and containerized execution without forcing users to become container orchestrators.

\paragraph{Namespace semantics.}
The resulting isolation model is intentionally more relaxed than a full per-rank container sandbox.
The single container is started per-node with host IPC, network, PID, UTS, and cgroup namespaces, as required by the HPC-oriented Podman configuration.
This level of reduced isolation is common to all HPC container implementations, and it's finalized at preserving low-level, advanced inter-process communication techniques, standard scheduler control, signal propagation, and a consistent user experience with the host sessions, while still delivering the containerized software environment.

Skybox uses a namespace-entry model instead of starting new container namespaces for each Slurm rank, using the \texttt{setns(2)} system call to join the single container's user and mount namespaces while inheriting all other namespaces from host.
In practice, this is the part of the container context that remains relevant for the HPC use-case: the rank sees the containerized filesystem and credential mapping, while remaining fully visible to Slurm and host-level process management.

Having all ranks on a node join the same container namespaces is more than just beneficial in terms of resources and startup time, it is a necessity when employing unprivileged user namespaces in distributed computing scenarios: some low-level shared memory approaches used by communication libraries (Cross-Memory Attach used by OpenMPI is an example) rely on system calls which cannot cross boundaries between sibling user namespaces. In other words, separate unprivileged containers on the same host cannot use common shared memory mechanisms for faster communication. Additionally, joining all node ranks in the same namespaces automatically solves the preservation of host file descriptors (which are used by PMI implementations like PMI2) into containerized processes, which would otherwise require explicit file descriptor manipulation and precise selection for propagation through OCI runtime options, making the implementation substantially more complex and fragile.

Finally, the choice of the \texttt{setns(2)} system call to join the namespaces prioritizes low-overhead attachment. As quantified in Table~\ref{tab:container-startup}, the \texttt{Join container namespaces} step incurs negligible overhead, effectively reducing the attach to under 1 ms per rank. This performance is achieved by bypassing the container-runtime layer; in contrast, a full \texttt{podman exec} invocation would require a runtime handshake costing approximately $120$ ms. By avoiding per-rank runtime execution Skybox delivers a lightweight attach path that preserves standard scheduler control an signal propagation while still providing the containerized software environment.

\paragraph{Podman configuration and what remains unchanged.}
Skybox does not modify Podman's engine behavior. Instead, it supplies Podman with per-step storage root directories, an HPC-specific configuration, and the Parallax-backed image store and mount program.

For each Slurm step, Podman is given a step-local \texttt{graphroot}, \texttt{runroot}, and \texttt{pidfile}; the node-individual container is then started with an HPC-specific Podman configuration module, the shared Parallax image store exposed as an \texttt{additionalimagestore}, and the Parallax mount program registered as the overlay mount path. EDF fields are translated into common Podman run arguments that are passed through the Podman call. In this sense, Skybox preserves Podman as the upstream OCI engine and only supplies the scheduler- and storage-specific context around it.

It is worthy of note that, as part of the HPC-specific configuration options, Podman is instructed to map the current user's identifiers on the host to the same values inside the container (i.e. use Podman's user namespaces "\texttt{keep-id}" mode). This is done to offer a more consistent and seamless user experience between host and container environments and accommodate workflow actions that might rely on particular values of user or group names or identifiers.

\paragraph{Parallax orchestration in the Slurm lifecycle.}
Skybox also orchestrates Parallax as part of the job lifecycle. On a warm start, when the requested image already exists in the shared read-only Parallax store, Skybox skips acquisition entirely and local task~0 on each node starts the helper container directly from that store. On a cold start, global task~0 first pulls the image into a temporary local Podman store, invokes Parallax to migrate it into the shared SquashFS-backed image store, removes the temporary local copy, and then signals completion through a shared synchronization file. Only after this shared-store phase has completed do nodes proceed with node-individual container startup. This separation between a job-wide image acquisition phase and a node-local helper startup phase is what allows Skybox to combine Podman with Parallax efficiently at scale. From a operational perspective, the cold start is a first-use artifact provisioning step that populates the shared image store, whereas subsequent production launches normally reuse that prepared artifact via the warm start logic. As such, Sarus Suite treats image acquisition as an occasional job-wide case rather than as a step to be done on every run.

\paragraph{Policy control, CDI, and hooks.}
Skybox exposes two layers of control. First, ordinary EDF fields specify image, mounts, devices, environment variables, working directory, entrypoint behavior, and writability. Second, reserved \texttt{com.sarus.*} EDF annotations act as control-plane overrides for site-managed settings such as Parallax paths, Podman module selection, temporary storage locations, and optional features. These reserved annotations are consumed by Skybox itself and stripped before the final Podman invocation; they are therefore policy inputs to the Sarus Suite control plane, not OCI annotations passed into the runtime environment.

By contrast, non-reserved EDF annotations are forwarded to Podman unchanged. This means Skybox can carry CDI-style device selectors and OCI-hook-related annotations without implementing device resolution or hook logic itself. The precise runtime ordering of CDI processing and OCI hook execution is therefore delegated to Podman and the OCI runtime, which keeps Skybox focused on scheduler integration rather than re-implementing container-runtime policy.

\paragraph{Privilege model, failure handling, and cleanup.}
Skybox is designed for rootless Podman-based execution and creates per-user runtime directories and per-step temporary storage with user-only permissions. The model therefore avoids privileged per-rank container launch while still allowing the site to control runtime behavior through Podman modules, CDI exposure, OCI hooks, and overall Sarus Suite configuration. At the same time, it should not be interpreted as a strong isolation model, favoring scheduler control and host-integrated execution over full namespace isolation.

Failure handling follows the same division of responsibilities, where rank creation, tracking, signaling, and accounting remain completely under Slurm control. This ensures that the integration allows ranks to execute inside a container yet remain under full control of the Slurm lifecycle semantics, signal propagation, and task/cgroup configurations. For its part, Skybox detects missing or invalid EDFs, incomplete site configuration, failed image pull or migration, container startup failure, and attach-time errors such as missing namespace handles or inaccessible shim state. At job teardown, the last local rank on a node stops the helper container, per-step temporary directories are removed, and the shared synchronization file is deleted by a single designated node.
The Parallax image itself is intentionally left in the shared store so that subsequent jobs can reuse the warm cache. The implementation is open source, and we defer lower-level synchronization and corner-case details to that code base; here we emphasize the execution model and its role in the Sarus Suite architecture.

\subsection{Performance Extensions}
Once Skybox has created the per-node container context and forwarded standard annotations and device requests into Podman, Performance Extensions provide the host-integration layer through which Sarus Suite applies launch-time HPC adaptations on top of a portable container image. Their implementation consists of site-managed CDI resource descriptions together with OCI hooks deployed on compute nodes. The hooks are implemented in Rust and built as static binaries, which keeps deployment simple and startup overhead low across different system environments.

Within this layer, the goal is the late binding of performance-critical environment details. CDI and hooks fulfill distinct but complementary roles. CDI statically describes how selected devices and other host-managed resources should be exposed inside the container, including resources needed to complete scheduler-aware execution. OCI hooks can access the live container filesystem and apply runtime logic to perform additional dynamic modifications required for HPC use, such as host-library smart injection, configuration-file mounts, and related filesystem changes. Together, these mechanisms allow system-specific software and resources to be introduced at container start without embedding them permanently into the image.

A key capability is the Precreate Container Edits hook (\texttt{pce\_hook}), which operates at Podman's \texttt{precreate} stage. At this point Podman passes the proposed OCI runtime specification to the hook, allowing it to modify that specification before container creation proceeds. In its current form, \texttt{pce\_hook} focuses on the two classes of change that matter most for HPC runtime adaptation: environment variables and mounts. This is sufficient to inject communication libraries, accelerator runtimes, configuration files, scheduler-related resources, temporary filesystems, and scratch or software mounts directly into the container specification.

Complementing the \texttt{precreate} stage is a \texttt{prestart} hook that performs dynamic library replacements. To ensure seamless integration of HPC-tuned host libraries, this hook inspects the container dynamic linker cache and bind-mounts host counterparts over matching entries, according to platform-specific configurations. This approach preserves the link integrity of containerized applications without having to delve into linker cache or hardcoded rpaths. The hook supports runtime ABI compatibility checks to control compatibility replacement conditions. The hook can also inject a new library into the container if no matching target exists, ensuring that critical HPC components are available at runtime.

The mechanisms described above are used together with capability selection driven by either annotations or CDI names. An annotation or device entry in the EDF can request, for example, NCCL-related network support, libfabric provider injection for MPI interoperability, GPU runtime services, or Slurm resource integration. The corresponding CDI descriptions and hook configuration then realize that request using the libraries, devices, mount layout, configuration files, and environment settings appropriate for the local system. In this way, the annotation retains a stable semantic meaning across platforms even though its realization remains system-specific.

Two additional \texttt{prestart} hooks complement this path. The \texttt{ldcache\_hook} refreshes the container's dynamic linker cache by executing \texttt{ldconfig}, ensuring that host libraries introduced by CDIs or other hooks are visible to the linker at application launch. The \texttt{mps\_hook} manages the start of NVIDIA's Multi-Process Service (MPS) when requested, providing a lightweight mechanism for enabling scheduler-selected GPU-sharing behavior without modifying the image or adding per-application startup logic.

Performance extensions make host integration an explicit run-time action, through which Sarus Suite turns declarative runtime requests into a concrete HPC job environment at container start.

\subsection{Parallax}
Parallax adds an image representation and distribution layer that makes OCI container deployment for Podman practical at HPC scale. Its role is not to change the user-facing container workflow, but to adapt Podman's storage to the realities of large, shared systems: synchronized multi-node launches, parallel filesystems, diskless or storage-constrained compute nodes, and the high metadata cost of repeatedly unpacking layered OCI images. Rather than having each node pull its own copy of an image, Parallax converts images once into a shared runtime representation that can be reused across the system.

Parallax is implemented in Go on top of the \texttt{containers/storage} libraries used by Podman. This keeps the implementation close to upstream container internals while reducing maintenance overhead and preserving compatibility with rootless Podman workflows. Conceptually, Parallax extends Podman's overlay-based image storage with a parallel-filesystem-aware representation and mount runtime logic, while leaving standard image retrieval and container lifecycle unchanged.

Parallax implements a transparent adaptation of Podman's image store for distributed HPC execution. After a standard OCI image has been pulled, Parallax migrates it into a compressed, immutable SquashFS representation located in a shared read-only image store on the parallel filesystem. This enables a ``pull once, run everywhere'' model: the image is prepared once, then reused by many jobs and many nodes without repeated pull-and-unpack activity at launch time.

This addresses two common scale problems in HPC container deployment. First, it avoids the network and registry pressure that would arise if many nodes or users repeatedly pulled large images independently. Second, it reduces the metadata overhead associated with layered OCI images stored as ordinary unpacked files on shared filesystems. By replacing repeated per-node pulls with a shared compressed image representation, Parallax improves startup behavior while reducing contention on both storage and network infrastructure. Crucially, this approach circumvents a known limitation in rootless mainstream Podman: standard OverlayFS storage backends are incompatible with parallel filesystems (e.g., Lustre, GPFS) because these file systems typically do not support user namespace~\cite{rootless-podman-nfs}. Without this support, the backing store cannot correctly translate container UIDs into host permissions required for rootless overlay operations.

Parallax fits into the container lifecycle in three stages: image preparation, runtime mounting, and cleanup.

\paragraph{Image preparation.}
Users interact with images through \texttt{sarusctl} and Skybox, which abstract the invocations of Parallax underneath. Once an OCI image has been pulled, Parallax can migrate it into the distributed HPC read-only image store. During this step, the image's layers are flattened and converted into a SquashFS file placed on the parallel filesystem. At the same time, Parallax updates the corresponding storage metadata so that the migrated image remains visible and usable through Podman's internal storage view mechanisms. This preserves a consistent image-store abstraction while replacing the default unpacked representation with one better suited to distributed HPC storage.

\paragraph{Runtime mounting.}
At container launch, Podman continues to use its normal overlay storage driver interface, but the image contents are provided through a custom mount program. This program exposes the SquashFS image as a single read-only lower layer through an unprivileged \texttt{squashfuse} mount and combines it with a writable upper layer using \texttt{fuse-overlayfs}, having the resulting merged view become the container's root filesystem. As mentioned in Section~\ref{subsubsec:skybox-slurm-integration}, Sarus Suite preserves the host user identity via Podman's \texttt{keep-id} option rather than mapping it to root. However, this creates a permission mismatch where the unprivileged user cannot write to root-owned image contents.
To resolve this, the Parallax mount program leverages \texttt{squashfuse\_ll}, the low-level FUSE interface which allows remapping all files inside the mounted filesystem to an arbitrary \texttt{uid:gid} pair. Consequently, when the SquashFS image is mounted, all files and directories appear as owned by the unprivileged user. This enables applications to create working files or temporary content in arbitrary locations without modifying the container image or requiring privileged operations. The ID mapping is performed by FUSE and exists only in the mounted filesystem view, leaving the file ownerships in the original image unaltered.

\paragraph{Cleanup.}
After the container terminates, Parallax's mount program performs the corresponding unmount and resource cleanup operations. This includes a background watcher that ensures mounted resources are released correctly after use, preventing stale mounts from accumulating during repeated job execution.

Parallax is designed to satisfy several requirements that are important in production HPC settings. First, it is compatible with parallel filesystems: the shared image store is read-only, centrally visible, and reusable across compute nodes. Second, it works with rootless container execution, avoiding the need for privileged per-user image deployment workflows. Third, it integrates transparently with Podman through existing storage and overlay extension interfaces rather than through modifications to Podman's core engine behavior. Finally, it provides explicit image-management operations, such as migration and removal, to keep the shared store administratively manageable and consistent.

Within Sarus Suite, Parallax is the component responsible for turning portable OCI images into an HPC-efficient runtime representation. Podman still handles standard image acquisition and container lifecycle operations; Parallax supplies the shared, parallel-filesystem-friendly storage format and mount path that allow those same images to be instantiated efficiently at scale. This separation of concerns is important to the overall architecture: standard OCI workflows are preserved, while the HPC-specific optimization is introduced only in the system layer responsible for large-scale deployment.

\section{Scaling experiments}

We evaluate whether Sarus Suite’s architecture delivers on three properties: scheduler-aware container execution that preserves distributed-application scaling, launch-time host integration that enables portable images to achieve production HPC communication performance, and scalable image access that enables shared-filesystem performance at large scale. To test these points, we use four workloads spanning communication-intensive distributed HPC applications, large-scale NCCL-based AI training, and metadata-dominated startup behavior. Collectively, the set covers Python and C++ software stacks together with OpenMPI, MPICH, and NCCL communication libraries in both strong- and weak-scaling regimes; Table~\ref{tab:workloads} summarizes the corresponding software configurations.

\begin{table}[t]
\centering
\caption{Workload configuration and architecture validation.}
\label{tab:workloads}
\small
\setlength{\tabcolsep}{3.5pt}
\begin{tabular}{lllll}
\toprule
Workload & Type & SW stack & Comm stack & GPU \\
\midrule
PyFR        & CFD          & Python    & OpenMPI      & CUDA 12.8 \\
SPH-EXA     & SPH          & C++20     & MPICH        & CUDA 12.8 \\
Megatron-LM & LLM training & Python    & NCCL         & CUDA 13.0 \\
Pynamic     & FS metadata  & C/Python  & OpenMPI      & --        \\
\midrule
\multicolumn{5}{l}{\textit{\footnotesize Validation Focus: PyFR/SPH-EXA (Scaling \& Host Integration),}} \\
\multicolumn{5}{l}{\textit{\footnotesize Megatron-LM (OCI \& WLM Integration), Pynamic (Storage integration).}} \\
\bottomrule
\end{tabular}
\end{table}

Our primary comparison is against Enroot+Pyxis, which serves as the production HPC container baseline on this class of systems.
Established practice and literature about HPC containers have demonstrated that containerization does not introduce an appreciable performance impact~\cite{charliecloud-2017,sarus-2019,containers-no-performance-impact-canopie-2019,keller2023containers-survey}, therefore comparing with a state-of-the-art HPC container solution is representative of general performance on a given system using the same application stacks.
We also include an unoptimized Podman configuration as a baseline to isolate the contribution of Sarus Suite’s integration layers, in particular scheduler-aware execution, host-network integration, and scalable image access. The remainder of this section first describes the common experimental setup and then shows how each workload validates a distinct part of the Sarus Suite architecture.

\subsection{Experimental setup}

Experiments were conducted on the Alps HPC research infrastructure at CSCS, a Cray EX system in Lugano, Switzerland. The evaluated partition consists of GH200 nodes\footnote{Although the detailed study focuses on Alps \texttt{GH200}, we also deployed the same stack on CPU-only AMD EPYC Rome nodes, AMD EPYC + NVIDIA A100 nodes, and Ubuntu~24.04 CI on \texttt{amd64}, indicating portability across architectures and Linux environments.}, each with four NVIDIA Grace Hopper Superchips connected through an HPE Slingshot 11 interconnect, running CrayOS 24.8.0 on Linux 5.14 under Slurm. This environment is representative of the kind of production system on which site-specific communication and accelerator integration are necessary for correct high-performance execution.

We compare three runtime configurations. The \textbf{Sarus Suite} configuration uses upstream Podman 5.7.1 together with Sarus components for scheduler-aware execution, host-capability injection, and scalable image access. The \textbf{Enroot+Pyxis} baseline uses Enroot 4.0.0 and Pyxis 0.20.0 with the production extensions described in~\cite{cruz2024containers}, including startup from SquashFS images on the parallel filesystem, host network-stack injection, and EDF-based customization. The \textbf{Unoptimized Podman} configuration serves only as a comparison baseline: because standard Podman is otherwise non-functional in this environment, we enabled via Sarus Suite only the minimum required functionality for execution while disabling non-essential optimizations. Across configurations, images were stored on a Lustre parallel filesystem backed by HDDs and accessed over Slingshot.

High-performance execution on Alps requires launch-time access to both MPI- and CUDA-related host resources. The Slingshot interconnect was exposed through an HPE-specific libfabric implementation (libfabric 1.22.0), and NCCL-based workloads used the AWS OFI NCCL plugin 1.17.2 bind-mounted from the host. Sarus Suite supplies these resources through CDI and OCI hooks, whereas Enroot+Pyxis provide corresponding functionality through Enroot hooks. The comparison therefore evaluates not only end-to-end application performance, but also the effectiveness of different host-integration mechanisms for supplying site-specific communication support on Alps.

For distributed application runs, we used four processes per node to match the four GPUs available on each GH200 node. Node counts increased in powers of two, with the scale range chosen per workload according to its practical scaling limit. Each Sarus Suite and Enroot+Pyxis data point reports the mean and standard deviation over 10 runs. Repeated runs at a given scale were collected on the same allocation, with the two runtimes interleaved to reduce bias from time-varying load on the shared network and filesystem. The unoptimized Podman measurements are included only for qualitative reference: at larger scale, its plain-file root-filesystem exposure generated enough metadata and I/O pressure that repetitions had to be reduced to limit disturbance on the production system.

Pynamic required a different repetition strategy because its behavior is dominated by startup-time filesystem effects rather than steady-state computation. Repeating runs within a single allocation would have biased later measurements through kernel file caching, so repetitions were collected on separate allocations to preserve uncached startup conditions. Since Pynamic uses MPI only for synchronization barriers, this allocation strategy does not materially affect interpretation of the result.

Relevant experimental artifacts, including configuration files, Containerfiles, EDFs, and associated scripts, are provided in~\cite{sarus-suite-cug26-artifacts}.

\subsection{PyFR}
\label{subsec:scaling-pyfr}

\begin{figure*}[ht]
    \centering
    \includegraphics[width=1\textwidth]{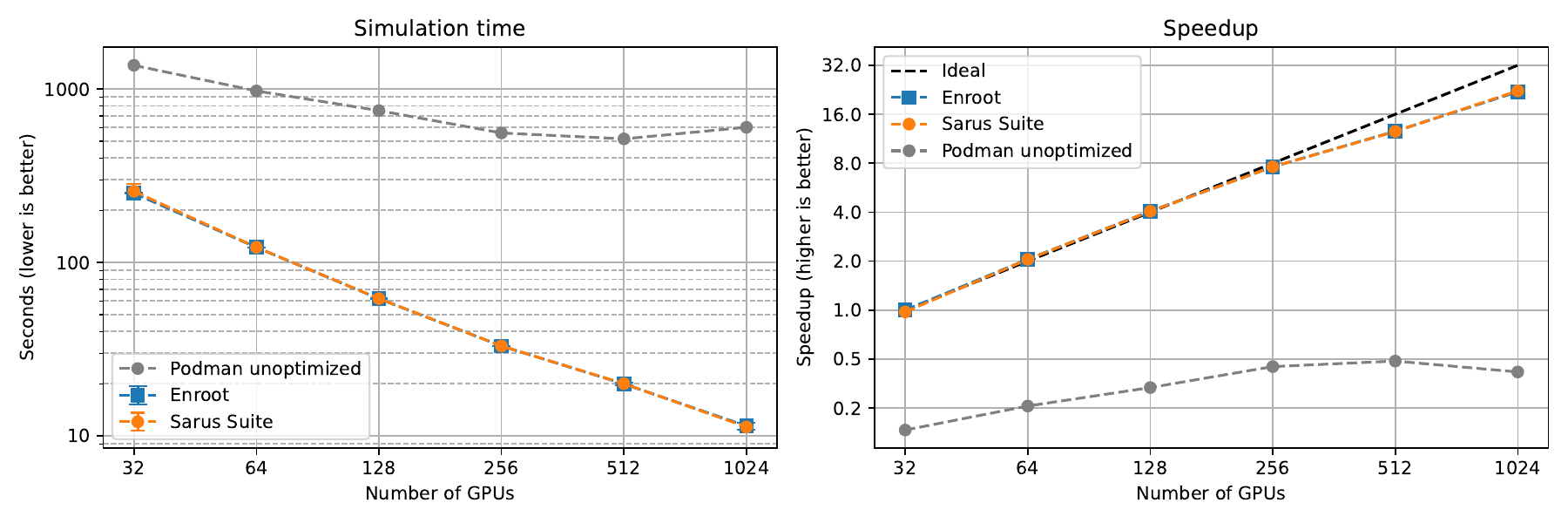}
    \caption{Performance and speedup results of the experiment with PyFR. Speedup values were computed using the performance average of each data point, taking the Enroot value at $32$ GPUs as baseline.}
    \label{fig:pyfr-results}
\end{figure*}

PyFR~\cite{pyfr-software} is a Python-based CFD code based on Flux Reconstruction and is known to scale well on GPU systems~\cite{pyfr-green-aviation-petascale}.
This test case evaluates the subsonic Taylor--Green vortex case from~\cite{pyfr-2.0.3-industrial-adoption-scale-resolving}, extending simulation time to $t_{end}=0.76~s$, disabling solution-file output to isolate simulation performance, and using a container with PyFR~2.1, OpenMPI~5.0.7, libfabric~1.22.0, and CUDA~12.8.1.

Simulations were run from 32 to 1024 GPUs (8 to 256 nodes), and we report net simulation time; the experiment is strong-scaling, with mesh and simulated time held constant across all runs.
Figure~\ref{fig:pyfr-results} shows that Sarus Suite and Enroot+Pyxis deliver nearly identical wall-clock time and speedup from 32 to 1024 GPUs, reaching $69.6$\% parallel efficiency at 1024 GPUs with low variability at all scales. For this OpenMPI-based strong-scaling workload, the close overlap indicates that Sarus Suite preserves efficient use of the Slingshot interconnect without introducing measurable scale-dependent overhead relative to the HPC container baseline.
By contrast, unoptimized Podman performs substantially worse and stops scaling meaningfully at larger node counts, showing that enhancing Podman scheduler-orchestrated one-container-per-node execution together with host-network integration is necessary for high performant MPI results.

\subsection{SPH-EXA}

\begin{figure*}[t]
    \centering
    \includegraphics[width=\textwidth]{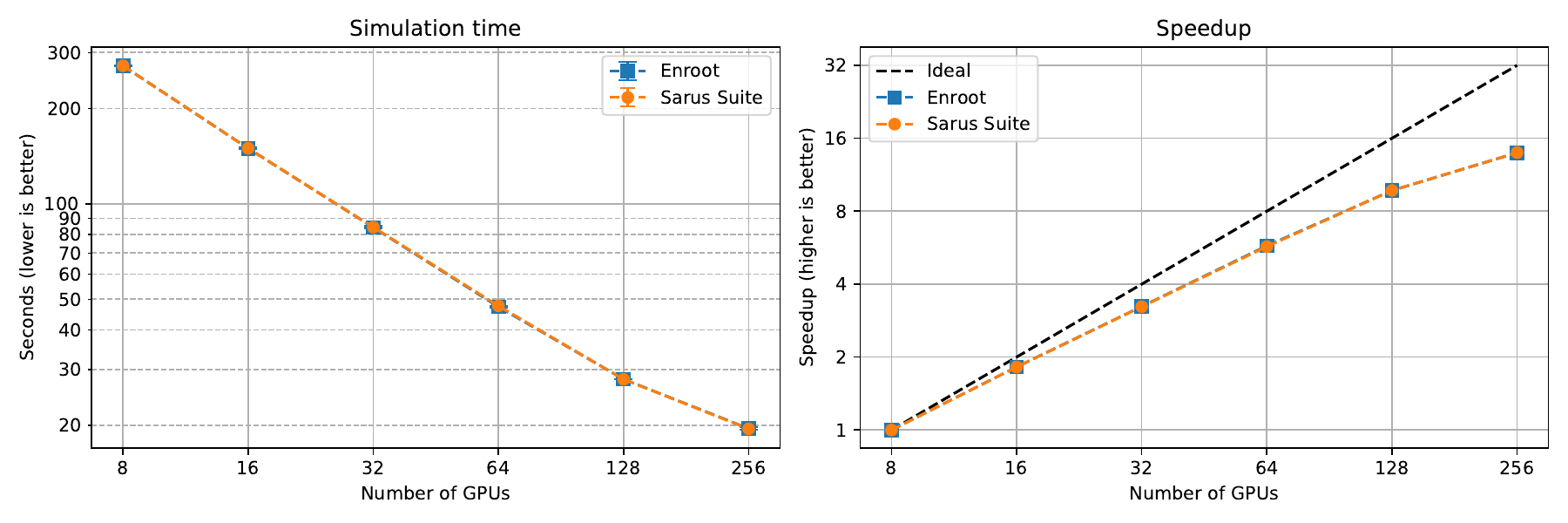}
    \caption{Performance and speedup results of the experiment with SPH-EXA. Speedup values were computed using the performance average of each data point, taking the Enroot value at $8$ GPUs as baseline.}
    \label{fig:sphexa-results}
\end{figure*}

SPH-EXA~\cite{sphexa-paper-2020} is a smoothed particle hydrodynamics simulation code designed for high performance and scalability on large-scale systems.
We use the built-in \texttt{evrard} test case, simulating the gravitational collapse of an isothermal cloud with $2144660433$ particles for $25$ time steps and $64$ OpenMP threads per MPI process, in a container with SPH-EXA~0.95, MPICH~4.3.1, libfabric~1.22.0, and CUDA~12.8.1.

Simulations were run from $8$ to $256$ GPUs (2 to 64 nodes), and we report total simulation-step execution time; this is a strong-scaling experiment, with particle count and time steps fixed across all runs.

Figure~\ref{fig:sphexa-results} shows that Sarus Suite and Enroot+Pyxis deliver nearly identical performance across all tested scales, and follow the same scaling trend up to $128$ GPUs, retaining $43.6$\% parallel efficiency at $256$ GPUs.

For this MPICH-based workload, the close agreement indicates that Sarus Suite preserves efficient host-network integration without introducing scale-dependent containerization overhead, and shows that the approach is not specific to OpenMPI-based applications.
We could not collect an unoptimized Podman baseline for SPH-EXA, because the application stalled during initialization without direct access to the Slingshot interconnect on Alps.

\subsection{Megatron-LM}
\label{subsec:scaling-megatron-lm}

\begin{figure}
\centering
\includegraphics[width=0.6\linewidth]{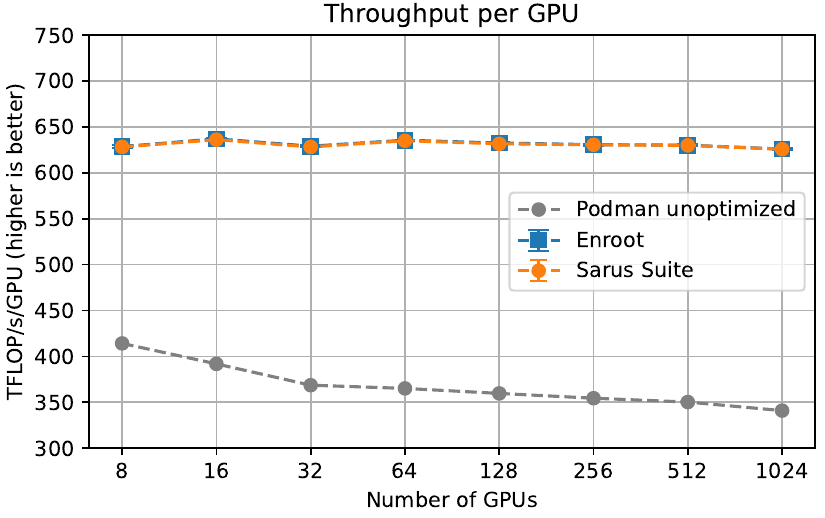}
\caption{Performance results of the experiment with Megatron-LM.}
\label{fig:megatron-results}
\end{figure}

Megatron-LM~\cite{megatron-lm-paper} is a transformer model training framework built on Megatron Core and PyTorch~\cite{pytorch-2-paper-2024}, representative of realistic large-scale NCCL-based AI training workloads.
We evaluate pre-training of the Llama 3 8B model~\cite{llama3herdmodels-grattafiori2024} using the bundled Megatron-LM example adapted to Slurm, with 4 GPUs per node, a Global Batch Size of $16\times$ the number of GPUs, mock data generation, 25 training iterations, and evaluation and checkpointing disabled; the container image was built from the PyTorch 25.11 NGC image with Megatron-LM and Megatron Core installed.
Training was run from 8 to 1024 GPUs (2 to 256 nodes), and we report per-GPU throughput averaged over the last 20 iterations of each run after discarding 5 warmup iterations; because the Global Batch Size scales with node count, this is a weak-scaling experiment.

Figure~\ref{fig:megatron-results} shows that Sarus Suite and Enroot+Pyxis deliver nearly identical per-GPU throughput at all tested scales, remaining close to 630~TFLOPS/s/GPU from 8 to 1024 GPUs, with minor throughput variations attributable to allocation topology and cluster network traffic.
For this NCCL-based workload, the close agreement shows that Sarus Suite performance extension preserves communication performance on the Slingshot interconnect while allowing the training job to run from an upstream NVIDIA NGC image rather than a site-specialized rebuild.
By contrast, unoptimized Podman delivers substantially lower throughput and degrades with scale, showing that OCI image portability alone is insufficient for distributed AI training unless runtime integration supplies the required site-optimized network components.

\subsection{Pynamic}

\begin{figure}
    \centering
    \includegraphics[width=0.6\linewidth]{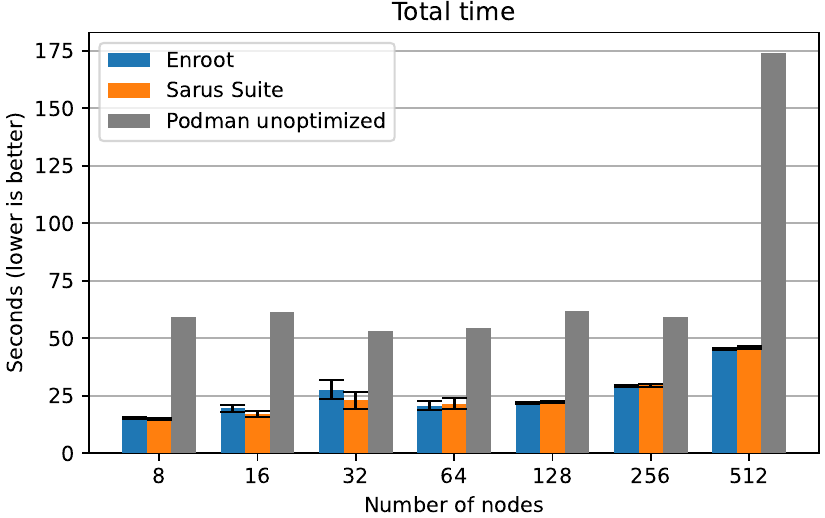}
    \caption{Overall results of the experiment with Pynamic.}
    \label{fig:pynamic-results}
\end{figure}

Modern scientific and AI/ML workloads often rely on Python, which generates a metadata-heavy access pattern at launch. The Python interpreter traverses directory trees to locate packages, open candidate files to validate imports, and invoke the dynamic linker for symbol resolution. In HPC environments where thousands of MPI ranks start simultaneously, this pattern saturates the metadata server, turning a seconds-long startup into minutes.

Pynamic~\cite{pynamic-paper,pynamic-software} is a synthetic benchmark designed to stress this specific behavior.
We use a containerized Pynamic installation comprising 495 files (280 Python modules and 215 utility libraries), each containing on average 1850 functions, to reproduce large-scale module import, traversal, and visit activity.
Runs were performed from 8 to 512 nodes, reporting the total time for startup, module-import, and module-visit phases. To ensure that measurements reflect actual filesystem performance rather than OS caching, we executed each repetition in a distinct Slurm allocation with epilog scripts configured to drop file caches between runs. This prevents the kernel from masking filesystem latency in subsequent repetitions, ensuring the reported times represent \textit{cold} filesystem access.

Figure~\ref{fig:pynamic-results} shows that Sarus Suite and Enroot+Pyxis deliver closely matching startup behavior across all tested scales, with total time remaining nearly flat up to 128 nodes and increasing only gradually thereafter.
For this metadata-dominated workload, the close agreement indicates that Sarus Suite preserves scalable launch-time filesystem behavior comparable to the production HPC container baseline.
By contrast, unoptimized Podman is consistently slower and degrades sharply at 512 nodes, showing that Parallax's SquashFS-based shared image representation is necessary to reduce metadata pressure and improved shared resource utilization relative to plain-file image layouts on parallel filesystem.

\subsection{Overall assessment}
The experiments support four main conclusions about the Sarus Suite approach. First, for distributed HPC applications, Podman complemented by Sarus Suite matches Enroot across both OpenMPI- and MPICH-based workloads. The PyFR and SPH-EXA results show that the combination of Skybox's one-container-per-node execution model and hook-based host integration avoids the scale-dependent penalties that would otherwise arise from inefficient runtime setup or incorrect exposure of the high-performance interconnect. In this sense, the Podman-based path is not merely functional, but competitive with a lightweight state-of-the-art HPC container runtime under strong-scaling conditions.

Second, the Megatron-LM experiment shows that this competitiveness can be achieved while preserving continuity with the mainstream container ecosystem. Starting from a vendor-distributed NVIDIA NGC image, Sarus Suite applies the system-specific communication integration at launch time, enabling NCCL to use the Alps Slingshot interconnect without rebuilding the image into a site-specialized variant. The resulting throughput matches Enroot across all tested scales, demonstrating that upstream OCI images can be carried much farther toward production HPC execution when the runtime provides the missing site-specific mechanisms.

Third, the Pynamic benchmark shows that the benefit of Sarus Suite is not limited to steady-state application performance. For metadata-dominated startup workloads, the Parallax image path yields launch behavior closely matching Enroot and substantially better than Podman backed by plain parallel-filesystem file trees. This indicates that scalable image representation and access are essential to making mainstream OCI engines practical on shared HPC storage at cluster scale.

The unoptimized Podman data, while used only as a baseline characterization, further clarify the contributions of Sarus Suite. Without scheduler-aware execution, performance-critical host integration, and filesystem-aware image handling, Podman exhibits severe penalties in both distributed communication and metadata-dominated startup. The combined evidence therefore supports the main architectural claim of this work: a mainstream container engine can be made HPC-grade without modification to the engine itself, so long as scheduler semantics, host capability injection, and scalable image access are provided through explicit, site-controlled integration layers.

\begin{table*}[!ht]
\centering
\caption{Per-node container startup time in the Slurm \texttt{slurmstepd} phase in seconds (mean $\pm$ std. dev.). Skybox is reported as an aggregated decomposition, with a detail on the per-task namespace-join. Enroot+Pyxis is reported for the corresponding total.}
\label{tab:container-startup}
\resizebox{\textwidth}{!}{%
\begin{tabular}{@{} l l r r r r @{}}
\toprule
\textbf{Stack} & \textbf{Measurement} &
\textbf{Ubuntu} &
\textbf{PyFR (No ext.)} &
\textbf{PyFR} &
\textbf{Megatron-LM} \\
\midrule
\multirow{4}{*}{Skybox}
& Podman-mediated startup
& 0.973 $\pm$ 0.056
& 1.691 $\pm$ 0.178
& 5.111 $\pm$ 0.403
& 9.680 $\pm$ 0.598 \\
& Runtime preparation
& 0.058 $\pm$ 0.008
& 0.058 $\pm$ 0.009
& 0.059 $\pm$ 0.007
& 0.059 $\pm$ 0.008 \\
& \quad of which: namespace join
& $\leq$ 0.001
& $\leq$ 0.001
& $\leq$ 0.001
& $\leq$ 0.001 \\
& \textbf{Total}
& \textbf{1.032 $\pm$ 0.062}
& \textbf{1.750 $\pm$ 0.184}
& \textbf{5.170 $\pm$ 0.402}
& \textbf{9.739 $\pm$ 0.599} \\
\midrule
Enroot+Pyxis
& Total
& 2.544 $\pm$ 0.017
& 3.211 $\pm$ 0.768
& 7.821 $\pm$ 0.782
& 11.338 $\pm$ 0.396 \\
\midrule
Relative
& Enroot+Pyxis / Skybox total
& 2.47$\times$
& 1.83$\times$
& 1.51$\times$
& 1.16$\times$ \\
\bottomrule
\end{tabular}
}
\end{table*}

\section{Container startup analysis}\label{sec:container-startup}

We analyze container startup timings for Skybox under Slurm, focusing on the per-node initialization cost relevant to large-scale production launches.

We consider four test cases of increasing complexity:
\textbf{Ubuntu}, the official Ubuntu 24.04 base image from Docker Hub, used as
a minimal baseline and as the base for the other test cases;
\textbf{PyFR with no extensions}, the PyFR image from
Section~\ref{subsec:scaling-pyfr}, representing a realistic application image
without HPC integrations;
\textbf{PyFR}, the same PyFR image but with the performance extensions used in
the scaling experiment enabled (support for NVIDIA GPUs and Slingshot network);
and \textbf{Megatron-LM}, the Megatron-LM image from
Section~\ref{subsec:scaling-megatron-lm}, representing a large ML/AI container
with all applicable performance extensions enabled, including support for the
AWS OFI NCCL plugin and an NGC-style entrypoint that adds measurable startup
work.

We instrumented Skybox to report elapsed times for functions executed during Slurm's \texttt{slurmstepd} stage, which runs on each compute node at the start of a job step. Each test case was executed with both Sarus Suite and Enroot+Pyxis for 10 repetitions on a single compute node, in order to isolate per-node startup latency from multi-node scaling effects and I/O contention. We then computed means and standard deviations for the measured times.

Container images were imported ahead of the measurements. This reflects the production workflow targeted by Sarus Suite: users typically validate the EDF and image interactively through \texttt{sarusctl}, including any required launch-time integrations, before submitting a large batch job, and this process also migrates the image into the Parallax shared store. Large Slurm launches therefore usually occur as warm starts, where Skybox reuses an already prepared shared image artifact rather than re-pulling and re-migrating the image. We therefore isolate the per-node startup cost of repeated production launches. Cold starts remain fully supported, but their runtime is dominated by registry/network conditions and one-time conversion into SquashFS, making them primarily an artifact provisioning cost rather than the representative cost of repeated large-scale execution through Sarus Suite.

Results are reported in Table~\ref{tab:container-startup}, together with a high-level decomposition of the Skybox per-node startup time. We observe that the dominant contribution to startup time comes from the Podman mediated startup phase (which includes image-availability check, performed through a Podman call; the container start, performed through \texttt{podman run}; and the wait for the container entrypoint to complete initialization), while the Skybox-specific runtime preparation remains small, and the per-task namespace join is negligible\footnote{In this analysis, considering that the startup of a Slurm job normally happens in the order of seconds, we consider negligible a value less than or equal to one millisecond.}.

The Podman-mediated startup time is only marginally affected by image size, whereas enabling HPC integrations through the performance extensions has a clear effect. Excluding the aggregated Podman-mediated startup, the overhead attributable to Skybox itself (captured as runtime preparation) remains less than 0.1\,s across all tested cases.

As described in Section~\ref{subsubsec:skybox-slurm-integration}, the image-availability check, the container start, and the wait for entrypoint completion are performed by only one rank on each compute node. Other co-located ranks wait for container readiness to join the container namespaces and complete their remaining setup steps (actions currently aggregated under \textit{Runtime preparation} measurement).

Table~\ref{tab:container-startup} also reports the corresponding total startup time of Enroot+Pyxis containers for direct runtime comparison. The HPC features enabled in Skybox through CDI and OCI hooks are matched in Enroot through its own hook mechanisms, providing feature parity for the comparison.

Skybox consistently outperforms Enroot+Pyxis in startup time across all test cases. While Skybox's own orchestration overhead remains negligible, when including the underlying Podman runtime, Skybox completes container initialization between 1.83$\times$ faster for PyFR without extensions and 1.51$\times$ faster for PyFR with extensions than Enroot+Pyxis, as reported in Table~\ref{tab:container-startup}. This advantage is most pronounced in baseline configurations, like the Ubuntu test case, where image complexity and entrypoint execution do not dominate startup time at 2.47$\times$ faster than Enroot+Pyxis. One possible contributor is the different implementation approach: Sarus Suite components are compiled binaries, with Podman implemented in Go, whereas Enroot relies heavily on Bash scripts for engine and hook execution. The present measurements do not isolate the source of the gap, but the result is consistent with lower orchestration overhead in the Sarus Suite stack.

These measurements show that, in the production-relevant warm start workflow, Skybox adds only a small and stable scheduler-aware orchestration overhead, while most of the remaining startup time is spent in runtime-mediated container creation and entrypoint execution. The comparison of total skybox time, which encompases runtime preparation and podman startup, between PyFR without extensions ($1.750$\,s) and PyFR ($5.170$\,s) isolates the added cost of enabling GPU and interconnect integrations, showing that the performance extensions introduce a real but bounded cost within Podman’s container-start step. With matched runtime features, Sarus Suite remains consistently faster than Enroot+Pyxis across all tested cases.

Relevant artifacts for the tests carried out in this section are also collected in~\cite{sarus-suite-cug26-artifacts}.

\section{Cloud-native workflows with Sarus Suite}

In this Section we discuss how the same Sarus Suite architecture also supports a qualitatively different type of workflow: cloud-native multi-container pods expressed through standard Kubernetes manifests. The focus here is therefore to present this feature and the architectural continuity with cloud-native capabilities.

Although HPC container workflows are often presented as single-container deployments, many contemporary scientific and machine-learning applications require richer execution models as those of cloud-native workflows with multiple cooperating containers, staged initialization, and shared transient storage. Sarus Suite addresses this need by enabling coordinated multi-container workflows through Podman’s support for standard Kubernetes YAML manifests. In this setting, the pod becomes a single coordinated execution unit in which multiple containers share a local runtime context, rather than being a collection of unrelated applications. To demonstrate this capability, we deploy a Ray-based~\cite{moritz2018ray,ray_software_2026} workload representative of task-parallel execution patterns relevant to reinforcement learning and large language model inference.

This approach is particularly attractive in HPC because it reuses a widely adopted declarative interface rather than introducing a new HPC custom abstraction. By leveraging Podman’s cloud-native functionality, Sarus Suite can interpret standard Kubernetes manifests to express architectural patterns such as staged initialization with init containers, transient intra-pod data exchange through ephemeral volumes, and persistent export of selected artifacts to the host filesystem. These mechanisms allow complex workflows to be described compactly while preserving the lightweight operational model expected in HPC environments.

\begin{figure}[t]
\begin{minipage}{\linewidth}
\begin{lstlisting}[
caption={Minimal conceptual sketch of the Ray demo under \texttt{sarusctl run ray.yaml}: an init container prepares the workload in a shared volume, a Ray head and worker form the runtime, and a submitter executes the client.},
basicstyle=\ttfamily\scriptsize,
breaklines=true,
columns=fullflexible,
frame=single,
label={lst:ray-sketch},
numbers=left,
]
apiVersion: v1
kind: Pod
metadata:
  name: ray-demo
  annotations:
    com.hooks.cxi.enabled: "true"
spec:
  restartPolicy: Never
  volumes:
    - { name: workdir, emptyDir: {} }
    - name: results
      hostPath: { path: /host/results }

  initContainers:
    - name: prepare
      image: python:3.14-trixie
      command: ["sh","-lc","<write /work/demo.py>"]
      volumeMounts:
        - { name: workdir, mountPath: /work }

  containers:
    - name: ray-head
      image: rayproject/ray:2.54.0
      command: ["sh","-lc","ray start --head --num-cpus=0 --block"]
      volumeMounts:
        - { name: workdir, mountPath: /work }

    - name: ray-worker-cpu
      image: rayproject/ray:2.54.0
      command: ["sh","-lc",
        "ray start --address=127.0.0.1:6379 --num-cpus=2 --block"]

    - name: ray-worker-gpu
      image: rayproject/ray:2.54.0
      command: ["sh","-lc",
        "ray start --address=127.0.0.1:6379 --num-cpus=2 --num-gpus=1 --block"]
      resources:
        limits: { "nvidia.com/gpu=all": 1 }

    - name: submitter
      image: rayproject/ray:2.54.0
      command: ["sh","-lc","python /work/demo.py; cp /work/*.json /results/"]
      volumeMounts:
        - { name: workdir, mountPath: /work }
        - { name: results, mountPath: /results }
\end{lstlisting}
\end{minipage}
\end{figure}

Ray is a relevant framework for this case study because it provides a unified programming model for a wide range of machine-learning workloads, including batch inference, distributed model execution, and reinforcement learning. Its programming abstractions its flexible supporting task submission, distributed state management, and dynamic resource coordination under the same framework runtime layer. As a result, Ray is a good use case for evaluating whether Sarus Suite can support the multi-process, stateful, and role-specialized execution patterns common in modern data-science pipelines within a lightweight HPC environment.

Listing~\ref{lst:ray-sketch} makes this coordination pattern explicit by exposing the workflow in the same order in which the pod is defined. First, the \texttt{volumes} section defines two storage scopes: an \texttt{emptyDir} volume (\texttt{workdir}) for transient data that is shared across containers and a \texttt{hostPath} volume (\texttt{results}) for exporting selected outputs to the host filesystem. Next, the \texttt{initContainers} section handles initialization: the \texttt{prepare} container generates workload data under \texttt{/work} before the main containers start. Then four containers implement distinct runtime roles: \texttt{ray-head} starts the local Ray head node, two Ray worker containers (one CPU-only, one GPU-enabled) join it, and \texttt{submitter} executes the client script and copies the resulting artifact to the persistent results mount. In this way, Listing~\ref{lst:ray-sketch} illustrates the use of shared storage and multi-container coordination to implement a structured pipeline. To enable this capability, \texttt{sarusctl} accepts as input a valid Kubernetes YAML manifests and forwards it to \texttt{podman kube} for instantiation.

Although the workflow is defined using a standard Kubernetes Pod manifest and launched via \texttt{sarusctl} rather than the Slurm-integrated Skybox, it retains full access to Sarus Suite's HPC oriented runtime optimizations. Specifically, \texttt{sarusctl} enables image access through Parallax and host-resource integration via CDI and OCI hooks, alongside the site-specific Podman namespace and cgroup configurations necessary for performant HPC execution. Furthermore, specific mechanisms such as OCI annotations and CDI names, previously demonstrated in Section~\ref{subsec:edf} via EDFs, are idiomatic to Kubernetes YAML and readily available for use. Listing~\ref{lst:ray-sketch} illustrates this by applying a pod-level annotation to enable an optimized network hook for all containers (enabled in lines 5–6, used by containers \texttt{ray-head}, \texttt{ray-worker-cpu}, \texttt{ray-worker-gpu}, \texttt{submitter}) and utilizing the NVIDIA CDI to enable GPU access specifically for the \texttt{ray-worker-gpu} container (lines 33--38).

The pattern demonstrated is especially relevant to HPC and scientific computing workflows, where applications frequently depend on staged initialization, shared node-local filesystems, and tightly coordinated task execution. The example in Listing~\ref{lst:ray-sketch} shows that Sarus Suite can express such non-trivial multi-container workloads cleanly through Kubernetes-style pod semantics, while still benefiting from HPC-oriented capabilities such as scalable image distribution through Parallax and performance-oriented extensions exposed via OCI hooks and CDIs. In this sense, the Ray example is not just a containerized application demo, but a compact demonstration that Sarus Suite can support modern composable workflow structures without requiring a full external orchestration stack.

To our knowledge, prior HPC container solutions have not shown this combination of standard Kubernetes-manifest-driven multi-container workflows with HPC specific image staging and runtime host integrations under fully OCI-compatible container engine.

The manifest and sample result for this example are provided in \cite{sarus-suite-cug26-artifacts}.

\section{Related work}\label{sec:related-work}

Container support for HPC is now well established, and existing solutions span different parts of the design space. Early HPC-specific implementations such as Shifter~\cite{shifter-2015}, Singularity/Apptainer~\cite{singularity,apptainer-software}, and Charliecloud~\cite{charliecloud-2017} showed that containerized software can be deployed on production supercomputers at acceptable operational cost. However, these systems generally rely on HPC-specific runtimes, image formats, or execution models. More recent solutions move toward modularity and upstream-aligned approaches, as in Sarus~\cite{sarus-2019}, Enroot+Pyxis~\cite{enroot-software,pyxis-software}, and podman-hpc~\cite{podman-hpc-2022}, but they still differ substantially in how they realize HPC integration. In contrast to existing systems, Sarus Suite brings together three properties in a single architecture: scheduler-native execution orchestration, a mainstream OCI engine, and OCI-oriented HPC host integration.

Among current systems, Enroot+Pyxis is the closest reference for the scheduler-native execution pattern of Sarus Suite. Enroot is a lightweight unprivileged runtime written in Bash, with fast image import and per-user container handling built around SquashFS-backed images, but it does not follow the OCI runtime lifecycle. Pyxis complements Enroot through Slurm SPANK integration, providing automatic image import, host environment propagation, and a one-container-per-node model in which all tasks on a node join the same container. Container deployment is exposed primarily through Slurm command-line options, which aligns naturally with scheduler-driven HPC workflows. Relative to Sarus Suite, the main difference is architectural: Enroot+Pyxis combines a custom lightweight runtime with scheduler integration, whereas Sarus Suite relies on Podman as a mainstream OCI engine and augments it with complementary components for runtime declaration, host integration, and parallel-storage support.

Podman-hpc is the closest prior work in terms of pursuing an upstream-aligned Podman-based solution. Both podman-hpc and Sarus Suite seek to enable production HPC workflows on top of a mainstream Podman ecosystem rather than replacing it with a domain-specific engine. Podman-hpc demonstrated the feasibility of this approach through rootless operation, SquashFS-backed image execution, and wrapper-managed execution modes. However, its abstraction boundary remained centered on user-facing wrappers and per-node \texttt{exec-mode} orchestration, and the authors report operational challenges at scale, including shared-storage configuration-file contention and sensitivity to metadata-heavy startup workloads. Sarus Suite builds on the same goal but shifts the integration point: EDF makes runtime intent a reusable object, Skybox embeds container lifecycle directly into Slurm using an efficient per-rank attachment mechanism, and Parallax combines one-time image provisioning with warm-start reuse through a robust shared read-only image store. In this sense, whereas podman-hpc adapts Podman from outside through wrappers, Sarus Suite integrates Podman into a scheduler-native, declarative, and lifecycle-aware execution model.

Prior work on the Sarus engine provides the closest reference for the OCI-oriented host integration aspect of Sarus Suite. Sarus introduced a modular architecture grounded in OCI standards, enabling the use of upstream and OCI-compliant components for image import and conversion, container execution, and runtime extensions. It also demonstrated how OCI hooks, provided either by the project itself or by third parties, can expose HPC features at container launch and thereby uplift portable generic images to HPC performance. However, Sarus is a privileged setuid-root custom engine mainly designed for traditional batch jobs, with limited support for the OCI runtime lifecycle, interactive workflows, workload manager integration, and more recent cloud-native features. Sarus Suite carries forward the same integration philosophy, but relocates it into a Podman-based architecture.

In this sense, Sarus Suite can be viewed as an architectural synthesis of ideas explored in earlier separate works, brought together here to address, within a single design, the use of a mainstream OCI engine, full workload-manager execution-model integration, and OCI-oriented HPC host integration.

Shifter, Singularity/Apptainer, and Charliecloud remain important points of reference because they illustrate other successful design choices in HPC containerization: privileged or mixed-privilege runtimes, runtime-centric interfaces, single-file image workflows, and lightweight unprivileged host-integration models. These systems established many of the practical trade-offs between portability, site policy, and performance that still shape the field. Relative to this broader landscape, Sarus Suite occupies a distinct position: it does not introduce another HPC-specific container engine, but instead seeks to retain a mainstream OCI engine while adding scheduler-native orchestration and OCI-based HPC integration as first-class architectural elements.

\section{Conclusions}

Sarus Suite demonstrates that production HPC systems can combine software agility with scheduler control, scalability, and performance while remaining aligned with the mainstream cloud-native ecosystem. By keeping Podman unmodified and implementing HPC-specific functionality through scheduler integration, scalable image staging, and standards-based performance extensions, Sarus Suite delivers production-ready HPC capabilities without requiring an HPC-specific engine.

Experiments on a Cray EX GH200 system support this claim across communication-intensive HPC applications, large scale AI training, and metadata-heavy Python workloads. Sarus Suite matches the state-of-the-art Enroot+Pyxis baseline in application performance while delivering faster container startup in our measurements. At the same time, Sarus Suite preserves ecosystem continuity by allowing users to deploy portable upstream images, such as NVIDIA NGC containers, and to express richer cloud-native patterns, including multi-container workflows defined through Kubernetes manifests.

With Sarus Suite, we propose a practical direction for HPC containers: preserve standard, cloud-native container semantics and specialize only the HPC integration layer. This allows HPC sites to benefit directly from upstream container innovation while maintaining scheduler control, scalability, and performance without the cost of maintaining a divergent runtime.

\section*{Acknowledgments}
This work was supported by the Swiss State Secretariat for Education, Research, and Innovation (SERI) through the SwissTwins project.

The authors thank Dr. Peter Vincent for kindly sharing the test case used in the PyFR scaling experiment.

The authors employed AI tools (specifically Qwen 3.5, ChatGPT-5, and Grammarly) to help improve the readability and grammar of this manuscript. All ideas, data, and conclusions presented are original to the authors, who assume full responsibility for the content.

\bibliographystyle{ACM-Reference-Format}
\bibliography{references}

\appendix

\end{document}